# Block Free Optical Quantum Banyan Network based on Quantum State Fusion and Fission[*]

Zhu Chang-Hua (朱畅华)[†], Meng Yan-Hong（孟艳红）, Quan Dong-Xiao (权东晓),

Zhao Nan (赵楠), Pei Chang-Xing (裴昌幸)

State Key Laboratory of Integrated Services Networks, Xidian University, Xi'an 710071, China

**Abstract:** Optical switch fabric plays an important role in building multiple-user optical quantum communication networks. Owing to its self-routing property and low complexity, Banyan network is widely used for building switch fabric. While, there is no efficient way to remove internal blocking in Banyan network by classical way. Quantum state fusion, by which the two-dimensional internal quantum state of two photons could be combined into four-dimensional internal state of a single photon, makes it possible to solve this problem. In this paper, we convert the output mode of quantum state fusion from spatial-polarization mode to time-polarization mode. By combining modified quantum state fusion, quantum state fission with quantum Fredkin gate, we propose a practical scheme to build an optical quantum switch unit which is block free. The scheme can be extended to build more complex units, four of which are shown in this paper.

# 1. Introduction

Based on the principle of quantum mechanism and unique quantum mechanical effects (e. g. quantum entanglement), quantum information technology can enhance the transmission, processing and storage of information, and has been developing quickly. Single photon, for its low decoherence rate and easier manipulation, has been widely used in quantum information technology (e. g. quantum key distribution[1]) and quantum computing (e.g. quantum logical gate[2][3]). With the development of photon-based quantum key distribution system, several experimental multiple-user quantum communication networks were built[4]-[8]. In these networks, trusted relay were used to implement long distance quantum key distribution. The key factor is that relay should be trusted. In short distance quantum key distribution system, untrusted optical switch can be used to build multiple users network [4][7].

[*] This work was supported by the National Natural Science Foundation of China (No. 61372076 and No. 61301171), the 111 Project (No.B08038), the Fundamental Research Funds for the Central Universities (No. K5051201021), and the scholarship from China Scholarship Council (No. 201308615037).
[†] Corresponding author. chhzhu@xidian.edu.cn

In classical communication network, switch-based topology is widely applied. In optical fiber communication network, optical switch is used to build optical cross connect (OXC) by which different devices can be connected. For optical quantum, OXC can be also applied to build an end-to-end quantum channel for two quantum communication terminals. Various switch fabrics were proposed for OXC, e.g. Crossbar network, Banyan network, Beneš network, Clos network, et. al.[9] A higher performance OXC should be lower in block rate, smaller in insert loss and shorter in processing delay. Banyan network has self-routing property since a qubit can be routed by only knowing it's input and output addresses, as shown in Fig.1. Each 2×2 switch can be either in cross state or in through state. A qubit is routed to the lower output of a switch at the $i^{th}$ stage if the $i^{th}$ most significant bit is '1' and to the upper output if it is '0'. Although Banyan network has simple structure and short processing delay, it also has a shortcoming that block exists. In classical optical network, it is a difficult problem except for adopting a more complex structure. While, there exists efficient method to overcome block in Banyan network by quantum characteristics, e.g. quantum superposition[10]. Several quantum switch schemes were proposed[11]-[14], but practical optical quantum circuit has not been proposed. Quantum state fusion and quantum state fission combining with quantum Fredkin gate may be one of efficient ways to remove block.

Quantum state fusion and quantum state fission are two processes proposed by Vitelli et. al. [15]. Quantum state fusion is defined as the physical process by which the internal quantum state of two particles (e.g. photons) is combined into four-dimensional internal state of a single particle (photon) [15]. For photons, the input photons may be in two-dimensional polarization mode, the output photon may be in four-dimensional spatial-polarization mode.

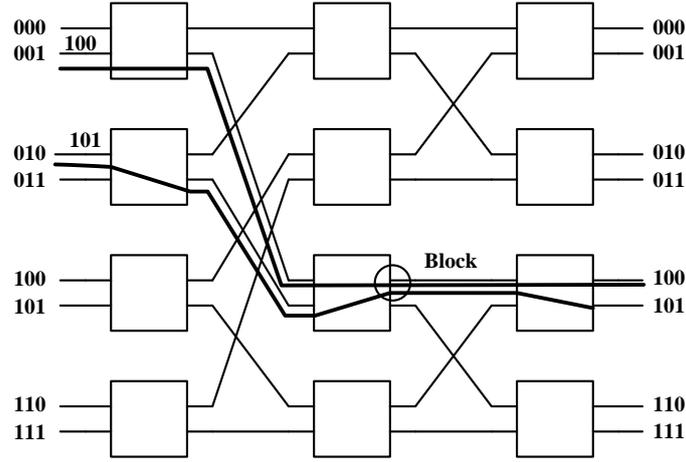

**Fig. 1 Block in banyan network**

An inverse process, quantum state fission, is defined as the physical process by which a higher-dimensional quantum state of a particle is split into two (or more) particles. By the quantum state fusion process, the blocking in Fig.1 can be efficiently removed. In the output of the second stage, a state fusion process could be applied to implement information fusion. Then in the third stage, a state fission process should be used to split them again. Although this is a good solution, the spatial-polarization mode still needs two output physical channels. We design a state conversion circuit to convert the spatial-polarization mode into time-polarization mode. The base states in time mode are one time instant and another time instant which is separated by a constant interval with the first one. This time interval is larger than the full-power bandwidth of optical pulse and dead time of single photon detector.

In this paper, we use this modified quantum state fusion scheme, quantum state fission scheme and quantum Fredkin gate to build block free optical quantum switch units. The rest of the paper is organized as follows: in section 2, a heralded linear optical quantum Fredkin gate is presented. In section 3, quantum state fusion scheme with time-polarization mode as output is given. A quantum state fission scheme with state conversion circuit is also given. In section 4, a block free optical quantum switch unit with two two-dimensional polarization states as inputs is proposed. Four types of switch fabric units are also proposed for four different scenarios with fused state (or states) as input. In section 5, we conclude the paper.

## 2. Heralded linear optical quantum Fredkin gate

Quantum Fredkin gate is also named as controlled swap gate, that is, if the control bit is in state $|1\rangle$, then the two target qubits swap their states and otherwise, they remain in their initial states if the control bit is in state $|0\rangle$. Quantum Fredkin gate can be built by three CNOT gates[16]. There are two types of optical methods to implement Quantum Fredkin gate. One is linear optical method by which the Fredkin gate is built by linear optical components[17]; the other is nonlinear optical method by which nonlinear effect of optical components is used[18][19]. In linear optical quantum information processing, qubit could be implemented by the photon with special polarization direction. The schemes and experiments with linear optical method are developed quickly. Gong et. al. constructed a heralded Fredkin gate by using four CNOT gates and linear optical components[17], they also simplified the methods to post-selection ones which operate in the coincidence basis and required ancillary photons in maximally entangled state.

Here, the state $|0\rangle$ is prepared by a photon with horizontal polarization, denoted by $|H\rangle$ and the state $|1\rangle$ is prepared by a photon with vertical polarization, denoted by $|V\rangle$. The operation for optical quantum bit is converted to the operation of polarization of photon. For a Fredkin gate, the two input quantum bits are $|\psi\rangle_{a_1} = \beta_{1h}|H\rangle_{a_1} + \beta_{1v}|V\rangle_{a_1}$ and $|\psi\rangle_{a_2} = \beta_{2h}|H\rangle_{a_2} + \beta_{2v}|V\rangle_{a_2}$, where $\beta_{1h}$, $\beta_{1v}$, $\beta_{2h}$ and $\beta_{2v}$ are arbitrary complexes, and $|\beta_{1h}|^2 + |\beta_{1v}|^2 = 1$, $|\beta_{2h}|^2 + |\beta_{2v}|^2 = 1$, $a_1$ and $a_2$ denote different space modes, as shown in Fig.2[17]. So the joint state of the two input photons is

$$|\psi_{in}\rangle = |\psi\rangle_{a_1}|\psi\rangle_{a_2} \\ = \beta_{1h}\beta_{2h}|H\rangle_{a_1}|H\rangle_{a_2} + \beta_{1h}\beta_{2v}|H\rangle_{a_1}|V\rangle_{a_2} + \beta_{1v}\beta_{2h}|V\rangle_{a_1}|H\rangle_{a_2} + \beta_{1v}\beta_{2v}|V\rangle_{a_1}|V\rangle_{a_2} \qquad (1)$$

Quantum Fredkin gate shown in Fig.2 can be in "Cross" state or "Through" state controlled by the control qubit $|c\rangle_{in}$. In "Cross" state the two input qubits are swapped, the qubit $|\psi\rangle_{b1} = |\psi\rangle_{a2}$ and the qubit $|\psi\rangle_{b2} = |\psi\rangle_{a1}$. In "Through" state the two input qubits are

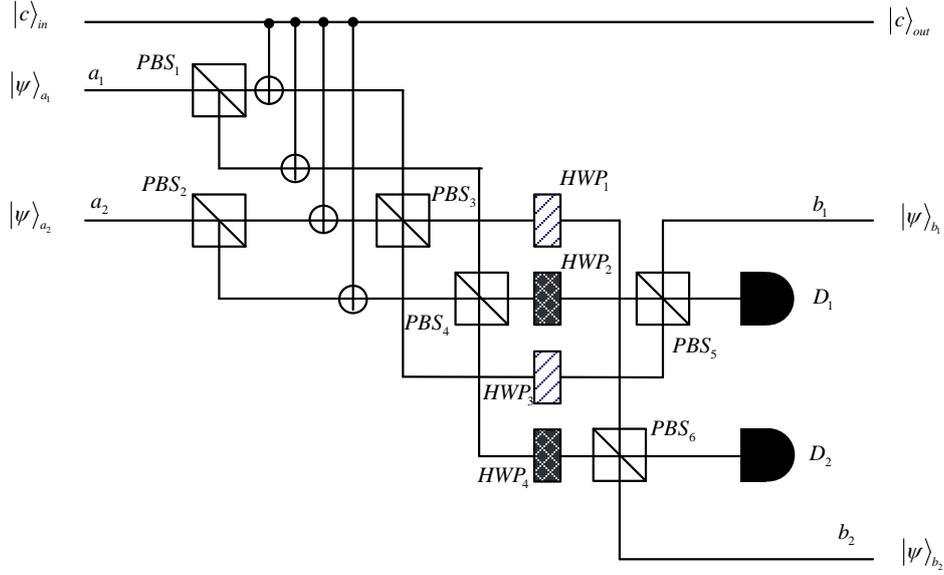

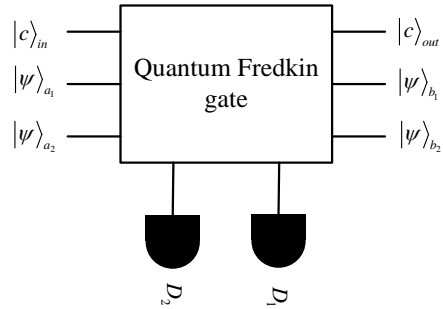

**Fig.2. A linear optical implementation of quantum Fredkin gate[17].**
**(a)quantum circuit. (b) a simple model in which only inputs and outputs**
**are given.**

passed through directly, the qubit $|\psi\rangle_{b_1} = |\psi\rangle_{a_1}$ and the qubit $|\psi\rangle_{b_2} = |\psi\rangle_{a_2}$. In Fig.2, $D_1$ and $D_2$ are two single photon detectors (SPD). CNOT gate can make the polarization direction of the photon rotate $90^0$. $HWP_1$ and $HWP_3$ are 22.5°-tilted half-wave plates(HWP), which act as Hadamard gates ($|H\rangle \to (|H\rangle+|V\rangle)/\sqrt{2}$, $|V\rangle \to (|H\rangle-|V\rangle)/\sqrt{2}$). $HWP_2$ and $HWP_4$ are 67.5°-tilted half-wave plates which perform the transformation: $|H\rangle \to (|V\rangle-|H\rangle)/\sqrt{2}$ and $|V\rangle \to (|H\rangle+|V\rangle)/\sqrt{2}$. Polarization beamsplitter (PBS) works as follows: the horizontally polarized photon is passed through, while the vertically polarized

photon is reflected. If $|c\rangle_{in} = |1\rangle = |V\rangle$, and no photon arrives at $D_1$ and $D_2$, the output state is

$$|\psi_{out}\rangle = \beta_{1h}\beta_{2h}|H\rangle_{b_2}|H\rangle_{b_1} + \beta_{1h}\beta_{2v}|H\rangle_{b_2}|V\rangle_{b_1} + \beta_{1v}\beta_{2h}|V\rangle_{b_2}|H\rangle_{b_1} + \beta_{1v}\beta_{2v}|V\rangle_{b_2}|V\rangle_{b_1}. \tag{2}$$

In this case, the quantum Fredkin gate is in "Cross" state. Otherwise, if $|c\rangle_{in} = |0\rangle = |H\rangle$ and no photon arrives at $D_1$ and $D_2$, the quantum Fredkin gate is in "Through" state. If the quantum efficiency $\eta$ of SPD is 100% and the dark count isn't taken into account, the success rate of the quantum Fredkin gate is $0.25$.

A simple model of quantum Fredkin gate is shown in Fig. 2(b) in which only the inputs and outputs are given.

## 3. Quantum state fusion and fission

In this section, we introduce the schemes of quantum state fusion and state fission proposed by Vitelli et. al. [15] and our modification for applying to optical quantum switch unit.

### 3.1 Scheme of quantum state fusion

Vitelli et. al. proposed a schematic of quantum state fusion[15], in which the two-dimensional quantum qubits of two input photons can be combined into a single output photon within a four-dimensional quantum space. Based on this scheme, we add a state conversion circuit and a control qubit $|c\rangle_{d_{in}}$, as shown in Fig.3. In Fig.3, the half wave plates are all oriented at 22.5° with respect to the H direction except for $HWP_6$ which is oriented at -22.5°. Let $|\psi\rangle_{a_3} = \beta_{3h}|H\rangle_{a_3} + \beta_{3v}|V\rangle_{a_3}$ and $|\psi\rangle_{a_4} = \beta_{4h}|H\rangle_{a_4} + \beta_{4v}|V\rangle_{a_4}$ denote two input states, where $\beta_{3h}$, $\beta_{3v}$, $\beta_{4h}$ and $\beta_{4v}$ are arbitrary complexes, and $|\beta_{3h}|^2 + |\beta_{3v}|^2 = 1$, $|\beta_{4h}|^2 + |\beta_{4v}|^2 = 1$. The ancillary photon enters the setup in $|H\rangle$ state, that is $|a\rangle = |H\rangle$. So the joint input state is

$$|\psi\rangle_{in} = \left(\beta_{3h}\beta_{4h}|H\rangle_{a_3}|H\rangle_{a_4} + \beta_{3h}\beta_{4v}|H\rangle_{a_3}|V\rangle_{a_4} + \beta_{3v}\beta_{4h}|V\rangle_{a_3}|H\rangle_{a_4} + \beta_{3v}\beta_{4v}|V\rangle_{a_3}|V\rangle_{a_4}\right)|H\rangle_a|c\rangle_{d_{in}}. \tag{3}$$

The quantum state fusion is successful if there is one and only one photon at each H-channel of $PBS_{11}$ and $PBS_{12}$, no photon at each V-channel of $PBS_{11}$ and $PBS_{12}$, one and only one photon at $b_3'$ and $b_4'$. The detailed working process of quantum state fusion is shown in Ref. [15] and its supplementary materials. After quantum state fusion, the spatial-polarization state at $b_3'$ and $b_4'$ is

$$|\psi\rangle_{b_{43}'} = \beta_{4h}\beta_{3h}|H\rangle_{b_4'} + \beta_{4h}\beta_{3v}|V\rangle_{b_4'} + \beta_{4v}\beta_{3h}|H\rangle_{b_3'} + \beta_{4v}\beta_{3v}|V\rangle_{b_3'}. \tag{4}$$

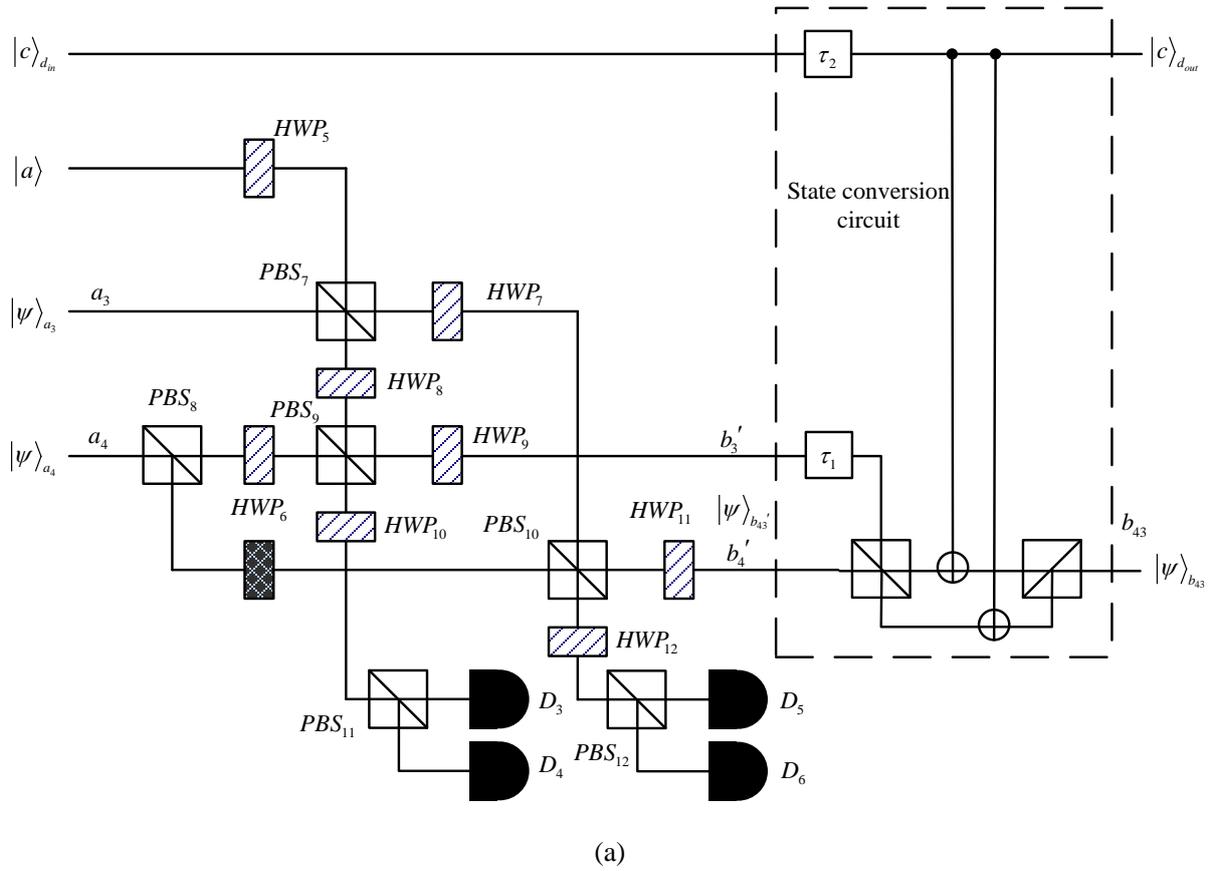

(a)

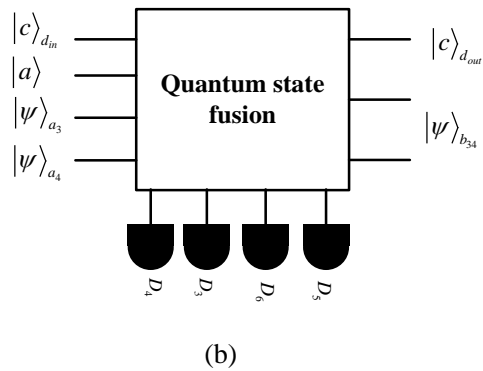

(b)

**Fig.3 Quantum state fusion. (a)quantum circuit. (b)a simple model in which only inputs and outputs are given.**

In Fig.3(a), state conversion circuit aims at converting the spatial-polarization mode to time-polarization mode. Here the state of a photon is in a four-dimensional Hilbert space obtained by time-polarization mode means that the photon may occur at one of the two time instants and may be in one of the two polarization directions. There are two time delay components with delay $\tau_1$ and $\tau_2$, two CNOT gates and two PBS components in the conversion circuit. The spatial modes at $b_3'$ and $b_4'$ are combined to one path (spatial) mode at $b_{43}$ based on the two CNOT gates which perform NOT operation only on the photons from $b_3'$ by adjusting the values of $\tau_1$ and $\tau_2$ properly.

A simple model is shown as Fig. 3(b) in which only the inputs and outputs are shown.

**3.2 Scheme of quantum state fission**

Vitelli et. al. also proposed an inverse process of quantum state fusion, quantum state fission, in which the four-dimensional quantum state of a single photon is split into two two-dimensional states of two photons, each carrying a qubit[15]. We add a state conversion circuit on their scheme, as shown in Fig.4. There are four input states: one fused state $|\psi\rangle_{a_{56}}$, one control state $|c\rangle_{d_{in}}$ under which state conversion circuit performs, and two ancillary states $|a\rangle$ and $|b\rangle$ which are in $|H\rangle$ at the inputs. In Fig.4, the half wave plates are all oriented at 22.5° with respect to the H direction, acting as Hadamard gates, except for $HWP_{23}$ and $HWP_{24}$ which are oriented at 45° so as to be equivalent to a NOT gate swapping states $|H\rangle$ and $|V\rangle$.

The detailed working process can be obtained in the supplementary materials of Ref.[15]. The state conversion circuit added here aims at converting time-polarization mode to spatial-polarization mode. The CNOT gates work only on the photon at the second time mode. After state conversion, the spatial-polarization state at $a_5'$ and $a_6'$ is

$$|\psi\rangle_{a_{56}'} = \beta_{5h}\beta_{6h}|H\rangle_{a_5'} + \beta_{5h}\beta_{6v}|V\rangle_{a_5'} + \beta_{5v}\beta_{6h}|H\rangle_{a_6'} + \beta_{5v}\beta_{6v}|V\rangle_{a_6'}$$

The quantum fission is successful if there is one and only one photon in the H-channel of $PBS_{20}$, no photon in the V-channel of $PBS_{20}$, one and only one photon in $b_5$, one and only one photon in $b_6$, no photon in $b_6'$. A simplified version of the scheme is shown in Fig.4 (b).

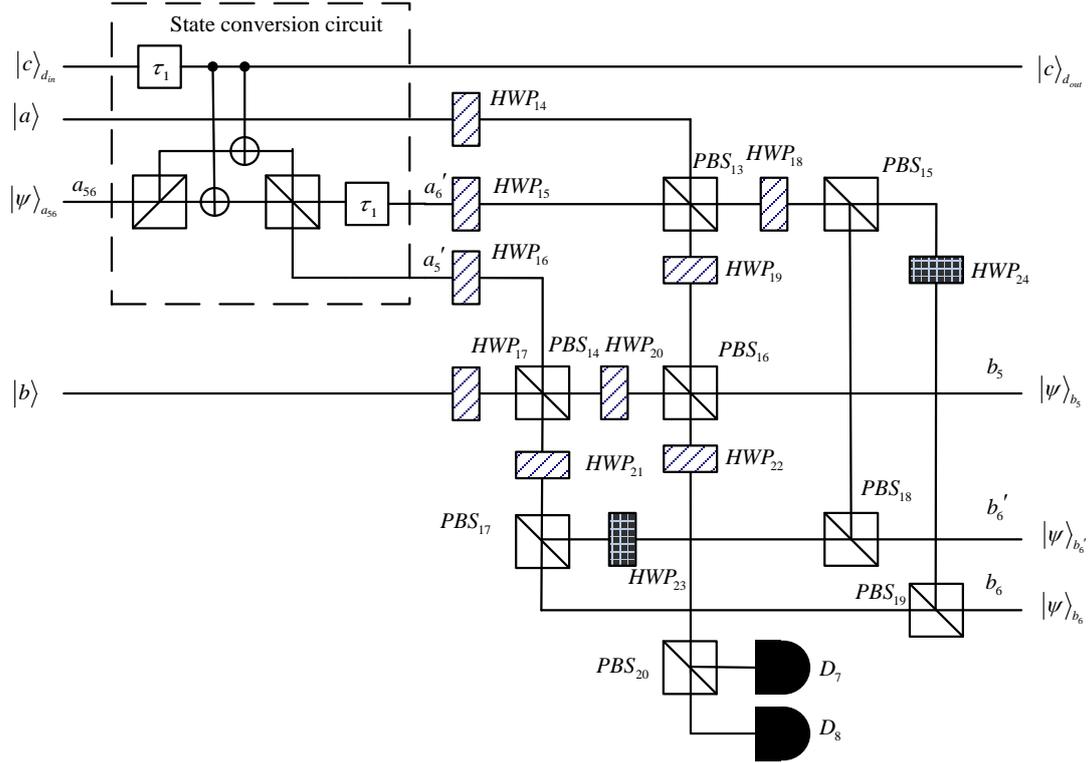

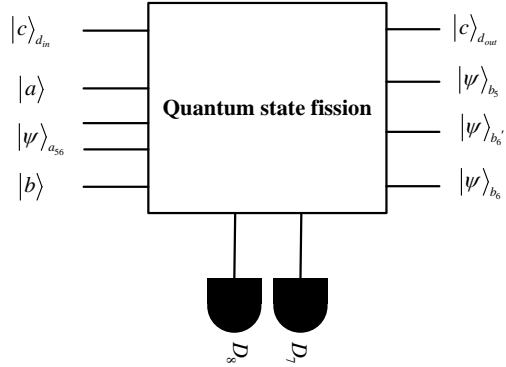

**Fig. 4. Quantum state fission. (a) quantum circuit. (b) a simple model in which only inputs and outputs are given.**

## 4. Block free optical quantum switch units

Based on the heralded quantum Fredkin gate in section 2, quantum state fusion and quantum state fission in section 3, we propose an optical quantum switch unit (OQSU) which is block free. Actually, the quantum state fusion schematic in section 3 can be directly applied to the scenario that two qubits which derive from two different inputs are switched to the same output at the same time. The two qubits can be fused and transmitted in the same quantum channel. Here, we propose a scheme in which the two input states may be fused or not.

### 4.1 Optical quantum switch unit (OQSU) with block free

The proposed optical quantum switch unit is shown in Fig. 5. There are two input information qubits at $a_7$ and $a_8$, $|\psi\rangle_{a_7}$ and $|\psi\rangle_{a_8}$, four control qubits $|c\rangle_{f_{in}}$, $|c\rangle_{F_{in}}$, $|c\rangle_{S_{in}}$ and $|c\rangle_{d_{in}}$, and one ancillary qubit $|a\rangle$. Let $|c\rangle_{f_{in}}$ denote the output competition. When $|c\rangle_{f_{in}} = |1\rangle$, it means that output competition exists and quantum state fusion works. When $|c\rangle_{f_{in}} = |0\rangle$, it means that there is no output competition and quantum Fredkin gate works. Let $|c\rangle_{F_{in}}$ denote the operation of quantum Fredkin gate which is in cross state when $|c\rangle_{F_{in}} = |1\rangle$, in through state when $|c\rangle_{F_{in}} = |0\rangle$. Let $|c\rangle_{S_{in}}$ denote the output port of the fused state which is $b_7$ when $|c\rangle_{S_{in}} = |1\rangle$, $b_8$ when $|c\rangle_{S_{in}} = |0\rangle$. Let $|c\rangle_{d_{in}}$ act as the control qubit of CNOT gate and is applied to combine the two spatial mode after delay of $\tau_2$. Let $|\phi\rangle$ denote the vacuum state. The input-output relation under different control qubits is shown in the Table 1.

In Fig. 5, path selection circuit consists of two CNOT gates and two polarization beam splitters. When $|c\rangle_{f_{in}} = |1\rangle$, the two input states $|\psi\rangle_{a_7}$ and $|\psi\rangle_{a_8}$ are transmitted to the output ports at $a_7''$ and $a_8''$. Otherwise, they are transmitted to the outputs at $a_7'$ and $a_8'$. The two states at $a_7'$ and $a_8'$ are transmitted to quantum Fredkin gate. They are swapped or not under the control of qubit $|c\rangle_{F_{in}}$. The two states at $a_7''$ and $a_8''$ are fused to a spatial-polarization mode. Then the spatial-polarization mode is converted into a time-polarizaiton mode by a converting circuit in which the time delay $\tau_1$ and $\tau_2$ are set properly. The output selection circuit is used to transmit the fused state to the expected output

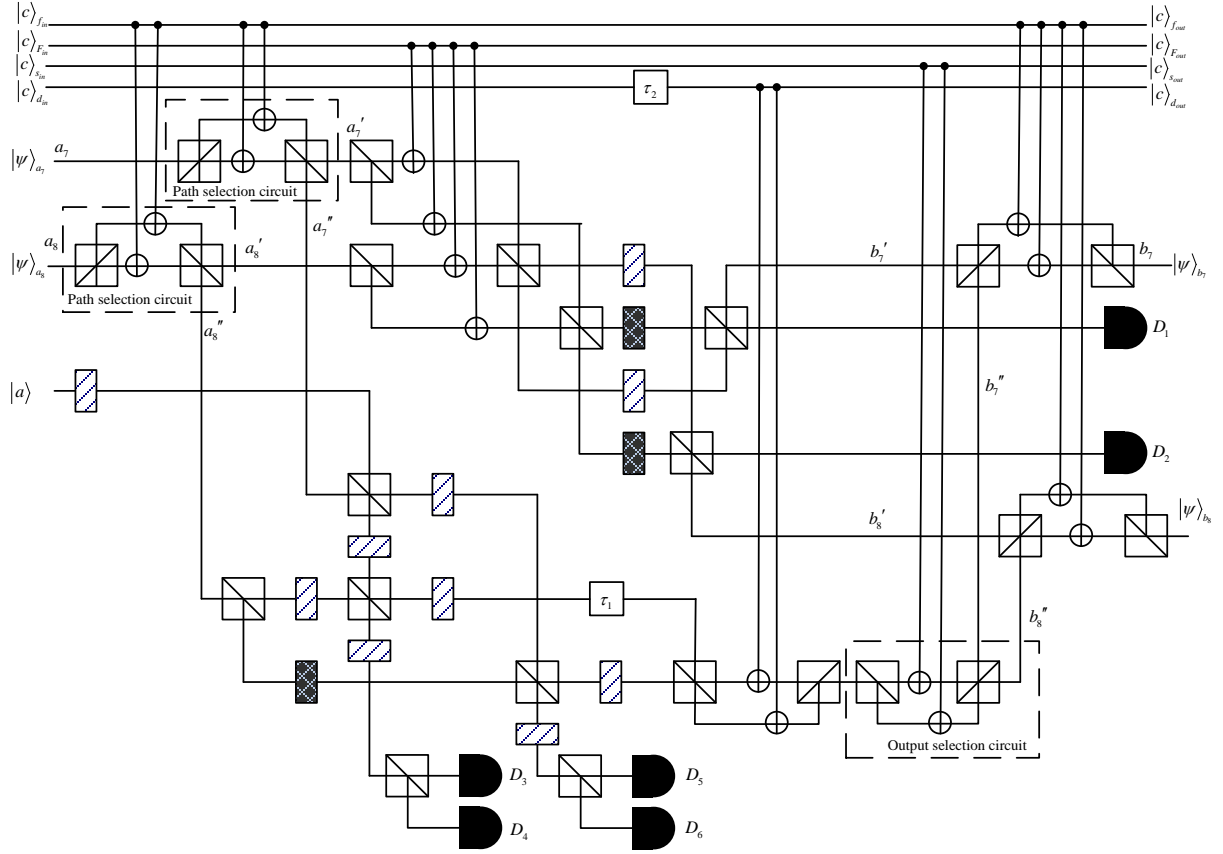

**Fig.5. An optical quantum switch unit with block free**

Table 1. The input-output relation under different control qubits

| $\|c\rangle_{f_{in}}$ | $\|c\rangle_{F_{in}}$ | $\|c\rangle_{s_{in}}$ | $\|\psi\rangle_{b_7}$ | $\|\psi\rangle_{b_8}$ | Condition |
|---|---|---|---|---|---|
| $\|0\rangle$ | $\|0\rangle$ | $\|0\rangle$ | $\|\psi\rangle_{a_7}$ | $\|\psi\rangle_{a_8}$ | |
| $\|0\rangle$ | $\|0\rangle$ | $\|1\rangle$ | $\|\psi\rangle_{a_7}$ | $\|\psi\rangle_{a_8}$ | Both $D_1$ and $D_2$ have |
| $\|0\rangle$ | $\|1\rangle$ | $\|0\rangle$ | $\|\psi\rangle_{a_8}$ | $\|\psi\rangle_{a_7}$ | no count. |
| $\|0\rangle$ | $\|1\rangle$ | $\|1\rangle$ | $\|\psi\rangle_{a_8}$ | $\|\psi\rangle_{a_7}$ | |
| $\|1\rangle$ | $\|0\rangle$ | $\|0\rangle$ | $\|\phi\rangle$ | $\|\psi\rangle_f$ | Both $D_3$ and $D_5$ detect one and only one photon. Both $D_4$ and $D_6$ have no count. There is only one photon in one of the two time modes at $b_7$ and $b_8$. |
| $\|1\rangle$ | $\|0\rangle$ | $\|1\rangle$ | $\|\psi\rangle_f$ | $\|\phi\rangle$ | |
| $\|1\rangle$ | $\|1\rangle$ | $\|0\rangle$ | $\|\phi\rangle$ | $\|\psi\rangle_f$ | |
| $\|1\rangle$ | $\|1\rangle$ | $\|1\rangle$ | $\|\psi\rangle_f$ | $\|\phi\rangle$ | |

under the control of qubit $|c\rangle_{s_{in}}$. So the two output states $|\psi\rangle_{b_7}$ and $|\psi\rangle_{b_8}$ can be a single state or fused state. The operation is performed according the value of the control qubits as shown in Table 1.

According to Ref. [15] and its supplementary materials, the state fusion shown in Fig. 3 is successful with a probability of 1/32. By a suitable feed-forward, the success probability can be raised to 1/8. The state fission shown in Fig. 4 is the same as the state fusion. Since the Fredkin gate has a 1/4 success rate at ideal condition, and we assume the value of the control qubit $|c\rangle_{f_{in}}$ is $|1\rangle$ or $|0\rangle$ with equal probability, the average success probability of the unit is 1/2×1/8+1/2×1/4=3/16.

Since heralded quantum Fredkin gate, heralded quantum fusion process and quantum fission process are used in each switch unit, a network will fail if any one of the units in the network could not work successfully. In this scenario, the quantum bits should be retransmitted. How to overcome this deficiency will be our further work.

**4.2 Optical quantum switch units with different input states**

Besides the quantum circuit in the Fig. 5, there are other block free switch units which consist of one fusion circuit or more fusion circuits, one quantum Fredkin gate or more Fredkin gates, on fission circuit or more fission circuits. Here are four examples, as shown in Fig.6.

In Fig. 6(a), the input state is a fused state, which is splitted into two single qubits by quantum state fission operation. Then a quantum Fredkin gate follows. This network can be used in a scenario in which a multiplexed optical quantum signal is demultiplexed and the two resulted separate qubits can be sent to two outputs as they require. In Fig. 6(b), the input states are one fused state and a single state, one of the state after fission is fused with the input single state. This network can be used in a scenario in which a multiplexed optical quantum signal is demultiplexed into two separate qubits, in which one qubit is transmitted to the output directly and the other is multiplexed with the input single photon signal. In Fig. 6(c), the input are one fused state and one single state. A vacuum state is inserted into the input state at next time slot in order to perform Fredkin operation. This network can be used in a scenario in which one multiplexed quantum signal is swapped with a single qubit or not. In Fig. 6(d), the input states are two fused states. They are fissed and fused again. This network can be used in a scenario in which two multiplexed quantum signals are demultiplexed and mutilplexed again after exchanging one of the two qubits. In these four scenarios, the switch units are successful if the counts of the detectors in the quantum circuits meet with the conditions shown in section 3 and there is one and only one photon in each output for non-fused sate and one and only one photon in each time instant for fused state. Note that the some scenarios are not taken into account, e.g. a fused state might compete with another input state for same output port. These scenarios should be further discussed.

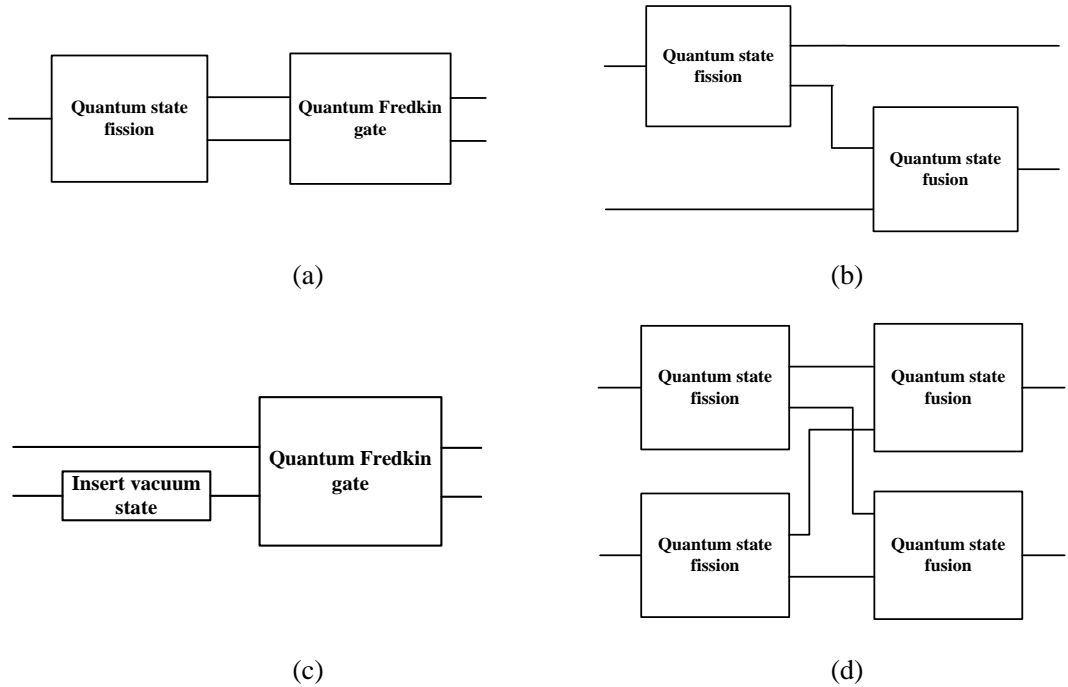

**Fig.6. Basic switch units based on the combinations of quantum state fusion, fission and Fredkin gate. (a) one fission and one Fredkin gate. (b) one fission and one fusion. (c) one Fredkin gate. (d) two fissions and two fusions.**

# 5 Conclusion

In this paper, we propose an optical quantum switch fabric with block free by using modified quantum state fusion, quantum state fission and quantum Fredkin gate. This switch fabric can be used to build a Banyan network with self-routing. We also propose schemes to build four types of switch fabric with block free. Since the photon loss in quantum circuit may decrease the fidelity of the switch fabric, we could improve its performance by using low-loss components and quantum nondemolition (QND) detectors.

Although quantum Fredkin gate, quantum fusion and fission processes are not be implemented with a success probability of 100%, the switch fabric proposed in this paper can be implemented without assisted classical control information and can be used to build all optical quantum information network. So, the side information leakage can be avoided.

With the development of integrated optics, large-scale optical quantum circuit or optical quantum computer may be built [20][21]. Our further works are to make experiment and to discuss how to implement it by integrated optics technology.